Hard Magnetic Materials _______________________________________________________________________________________

# Using of SPS technique for the production of high coercivity recycled SmCo$_5$ magnets prepared by HD process

Anas Eldosouky[1,2], Awais Ikram[2,3], Muhammad Farhan Mehmood[2,3], Xuan Xu[2,3], Sašo Šturm[2,3], Kristina Žužek Rožman[2,3], Irena Škulj[1]

[1] *Magneti Ljubljana, d.d., 1000 Ljubljana, Slovenia*
[2] *Jožef Stefan International Postgraduate School, 1000 Ljubljana, Slovenia*
[3] *Department for Nanostructured Materials, Jožef Stefan Institute, 1000 Ljubljana, Slovenia*

*Abstract*—Spark plasma sintering (SPS) technique is applied in combination with hydrogen decrepitation process for the recycling of SmCo$_5$ magnets. The SmCo$_5$ magnets for recycling were first decrepitated by hydrogen gas of a pressure of 4 bar for 3 hours to produce decrepitated powder. This powder was then used to prepare isotropic sintered magnets using the SPS technique, by sintering at 800-1000 °C for 1 minute. Full densification of the SPS-ed magnets was possible at a temperature of 1000 °C. The sample sintered at 900 °C showed the best internal coercivity ($_jH_c$) of higher than 1500 kA/m with high remanence ($B_r$) value of 0.47 T and energy product ($BH_{(max)}$) of 43.4 kJ/m$^3$. The properties of the SPS-ed sample sintered at 900 °C were compared with conventionally sintered (CVS-ed) sample prepared by using fresh SmCo$_5$ powder. The results showed the improvement of the magnetic properties of the SPS-ed sample in comparison to the CVS-ed sample at room temperature, and the possibility to use the SPS-ed sample at high temperature of 180 °C, where the sample showed good magnetic properties of $_jH_c$ of 1502 kA/m, $B_r$ of 0.44 T and $BH_{(max)}$ of 36.4 kJ/m$^3$. The microstructure and X-ray diffraction patterns of the SPS-ed and the CVS-ed samples were studied; where the samples showed to basically consist of SmCo$_5$ matrix phase with Sm$_2$Co$_7$ and Sm-oxides.

## I. INTRODUCTION

Permanent magnets are known to have many technical importances as they are used in a range of important applications as in motors, generators, sensors and hard disc drives. NdFeB and SmCo magnets dominate the market of permanent magnets production [Gutfleisch 2011]. SmCo magnets can be produced having compositions close to Sm$_2$Co$_{17}$ or SmCo$_5$. SmCo$_5$ offer very high anisotropy field of 31,840 kA/m, with excellent coercivity of a value of 2388 kA/m. Also, SmCo$_5$ magnets have very good temperature resistance; Curie temperature of 750 °C, reversible temperature coefficient of coercivity of -0.2 %/°C, and reversible temperature coefficient of remanence of -0.05%/°C [Pan 2014]. But, as SmCo$_5$ is composed of Sm and Co, and both elements are recently identified as elements of supply risk and technical importance; innovative recycling routes have to be defined for the magnet [European Commission 2017].

Hydrogen decrepitation (HD) is a process where the material absorbs hydrogen gas in its interstitial sites. This absorption causes volumetric expansion and the reduction of the material's particle size [Harris 1987]. The conditions needed for the decrepitation and the particle size produced depended on the material's chemical composition and microstructure. Recently, Walton et al. [2015] and Eldosouky and Škulj [2018] showed the possibility to use HD for the recycling of NdFeB and SmCo$_5$ magnets, by using conventional sintering (CVS) route for the densification of the decrepitated powder after milling. After the CVS of the decrepitated SmCo$_5$ powder, the microstructure of the original magnet, which consists of SmCo$_5$ matrix, and Sm$_2$Co$_7$ and Sm-oxides as randomly distributed phases, has been reformed.

Spark plasma sintering (SPS) is a process where a pulsed DC current passes through a die, which contains the powder of interest. The process is usually combined by an applied pressure. The mechanism of densification in SPS is different than CVS, where CVS consists of slow diffusion mechanisms such as grain boundary and volume diffusion. On the other hand, SPS is a rapid densification technique, where only the surface of the particle experience localized heating, with no observed contribution from the bulk of the particles [Diouf 2012]. Zhaohui et al. [2008] described the densification mechanism in SPS to consist of four steps; 1) powder activation and refining, 2) formation of sintering neck; these steps involve removing the surface oxide and heating the powder due to the voltage breakdown effect and formation of the neck by evaporation and condensation and diffusion, 3) growth of the

Corresponding authors: A. Eldosouky (anas.eldosouky@magneti.si) and A. Ikram (awais.ikram@ijs.si) contributed equally.
Digital Object Identifier: 10.1109/LMAG.2018.2831174.

sintering neck, and 4) plastic deformation densification, usually enhanced by the applied pressure. Due to the local heat generation, the sintering temperature required for the full densification of the powder is usually lower than the one required for the CVS, with much lower sintering time as low as minutes in comparison to days sometimes for CVS. There are many parameters which can be optimized in the SPS experiments, as, the pressure applied, the heating rate, the sintering temperature, the holding time and the partial pressure.

The SPS technique is now a well-established technique for the production of NdFeB magnets of high coercivity and excellent magnetic properties [Liu 2010, Mo 2007]. Moreover, the use of the SPS for the sintering of $SmCo_5$ magnets was shown to be an effective technique for the preparation of bulk nanocrystalline $SmCo_5$ or SmCo/Fe(Co) of good magnetic properties. Zhang et al. [2010] and Yue et al. [2011] used SPS for the preparation of bulk nanocrystalline $SmCo_5$ magnets. They showed that internal coercivity ($_jH_c$) value of 2273 kA/m and a remanence ($B_r$) value of 0.5 T can be achieved by using ball milled $SmCo_5$, where the sintering temperature was 700 °C at an applied pressure as high as 500 MPa. Fang et al. [2018] showed that, the use of wet-milling for the preparation of the $SmCo_5$ results in the oxidation of the powder during the processing, and the formation of $Sm_2Co_{17}$, with a decrease of $_jH_c$ of the magnets to a value as low as 280 kA/m. Saravanan et al. [2013] studied the magnetic properties and microstructure of $SmCo_5$/Fe nanocomposite bulk magnets at different temperatures between 700-850 °C, 75 MPa, where the highest coercivity of 654 kA/m was achieved at 800 °C, where the produced magnet had composition of mainly $Sm(Co,Fe)_5$ with other secondary phases such as $Sm_2(Co,Fe)_{17}$ and α-Fe(Co). Where, Saito and Nishio-Hamane [2018] showed that, decreasing the sintering temperature to 600 °C results in an enhancement in the coercivity to 980 kA/m. Lu et al. [2015] showed that, the $BH_{(max)}$ value for the SPS-ed $SmCo_5$ can be improved by the addition of NdFeB. $BH_{(max)}$ of 115 kJ/m$^3$ was obtained for SmCo/NdFeB (4:1), which is 50% higher than the single phase $SmCo_5$ magnets.

The aim of this study was to use a combination of HD and SPS route for the recycling of sintered $SmCo_5$ magnets and the production of isotropic $SmCo_5$ magnets at different sintering temperatures. The magnetic properties, microstructure, X-ray diffraction (XRD) patterns, density and oxygen content of the recycled magnets made by the SPS route are compared with those for industrially produced sintered magnet prepared by CVS route.

## II. EXPERIMENTAL PROCEDURES

Two $SmCo_5$ powders were used for this study. Freshly prepared powder of a particle size of 4-20 μm prepared by milling the as-cast alloy in a nitrogen atmosphere jet mill. The chemical composition of the powder was as follows: 36 wt% Sm, 63.8 wt% Co and 0.2 wt% O. The second type was the recycled powder prepared by HD of $SmCo_5$ sintered magnets. The magnets for recycling were collected out of industrial lines as they do not meet the shape specifications. The measured chemical composition of the magnet was 36.30 wt% Sm, 63.22 wt% Co, 0.37 wt% O. The decrepitation experiment was done in a rotating equipment as described by Eldosouky and Škulj [2018]. The conditions for the decrepitation reaction were 4 bar of hydrogen, and 250 rpm at room temperature for 3 hours. The produced powder after decrepitation was milled in a jet mill with nitrogen atmosphere to a particle size lower than 20 μm. 1 wt% of Sm metal was added for the milled powder to compensate for the oxygen in the starting material before recycling, and the oxidation during the decrepitation and further processing steps.

The milled decrepitated powder was sintered using the SPS route. Inside a glove box, 3 gm of the milled powder with the Sm addition was put in a 10 mm inner diameter graphite die. The die was then handled in air for less than 5 minutes to be put in the SPS machine, where in between, the powder inside the die was pressed by a hydraulic press to 5 bar. The SPS sintering parameters were 800-1000 °C sintering temperature (50 °C step), and heating rate of 50 °C/min. The sample was kept at the sintering temperature for 1 minute where the applied pressure was 80 MPa. After evacuation the SPS chamber to 0.05 mbar, partial pressure of 200 mbar Ar was kept during sintering to reduce the Sm evaporation. The sintering experiment was done by using SPS 25 series, model 615, Fuji electronic industrial CO., LTD. CVS route was used to produce $SmCo_5$ magnets made out of the fresh powder, without alignment similar to SPS experiments. A silicon mold of volume of 8 cm$^3$ was filled by the fresh powder to reach to a density of about 3 g/cm$^3$. The mold was then isostaticallly pressed at a pressure of 2500 bar. The pressed sample was then sintered at 1180 °C for 2 h and then heat treated for 3 h at 880 °C.

The microstructure analysis for the powder after decrepitation was carried by Helios NanoLab™ 650 scanning electron microscope, whereas, Jeol JSM-7600F scanning electron microscope was used for the sintered samples. PANalytical X'Pert pro multi-purpose diffractometer was used for the crystallography measurement. The demagnetization curves for the sintered samples were measured by Magnet-Physik's permagraph with maximum applied pressure of 1500 kA/m, after magnetization by 6 T field. The oxygen content was measured by Eltra ON 900, oxygen/nitrogen analyzer. The density was measured by Archimedes' principle.

## III. RESULTS AND DISCUSSION

Fig. 1 shows the secondary electron (SE) images for the produced powder after the decrepitation of $SmCo_5$ magnets. The decrepitated powder is of irregular shape with some sharp features and wide particle size distribution (~300 nm to ~200 μm). The powder is characterized with the presence of transgranular fractures in the $SmCo_5$ matrix phase, in addition to the intergranular fractures due to the hydrogen absorption by the $Sm_2Co_7$ phase as described by Harris [1987].
**FIG. 1 HERE**

Fig. 2 shows the demagnetization curves for the SPS-ed samples produced by sintering at 800 °C, 850 °C, 900 °C, 950 °C, and 1000 °C. Table 1 shows the magnetic properties, density and oxygen content for the magnets. The density has shown to increase with the sintering temperature from 6.86 g/cm$^3$ at 800 °C to 8.4 g/cm$^3$ at 1000 °C, which is about 100% of the theoretical density of the magnets. On the other hand, the oxygen content showed to decrease with increasing the sintering temperature, as the densification in SPS involves removing the surface oxide because of the voltage breakdown effect, which means increasing the sintering temperature, would lead to more efficient removing of the surface oxygen toward the full densification [Diouf 2012].

**FIG. 2 HERE**
**Table 1 HERE**

As expected from the relative relationship between the density and the sintering temperature; the $B_r$ and $BH_{(max)}$ of the samples showed to also increase with the sintering temperature to reach to a value of 0.51 T and 50.1 kJ/m$^3$, respectively, for the sample sintered at 1000 $^o$C. The sample sintered at 900 $^o$C showed a $B_r$ of 0.47 T which is about 50% of the value for the anisotropic SmCo$_5$ magnets produced industrially which have a $B_r$ of 0.80-0.96 T [Pan 2014].

For the samples sintered at lower than 900 $^o$C, due to the incomplete phases formation for those samples and the presence of high oxygen content and large pores percentage, the coercivity value was lower than the one for the industrial CVS SmCo$_5$ magnets, with $_jH_c$ of less than 1000 kA/m. Sintering at 900 $^o$C at the same conditions, has the effect of properly densifying the magnet to a reach to a density of 96% of the theoretical density. The measured $_jH_c$ value for this sample was higher than 1500 kA/m. Increasing the sintering temperature further showed to negatively affect the shape of the demagnetization curve, where the $_jH_c$ value was dropped to 1243 kA/m and 1164 kA/m for 950 $^o$C and 1000 $^o$C samples, respectively. The decrease of the $_jH_c$ upon increasing the sintering temperature is believed to be due to the excessive formation of the Sm$_2$Co$_7$, as it will be shown from the microstructure and XRD studies later, and smoothing out of the grain surface as described by Menushenkov [2006] for the decrease in the coercivity of the SmCo$_5$ magnets after heat treatment above 900 $^o$C for CVS magnets with Sm content of 36.4 wt%. Note that, the level of the carbon uptake from the graphite die during the SPS sintering is not enough to drastically change the magnetic properties of the produced magnets, as shown by Mackie et al. [2016] for Sm(Co, Fe, Cu, Zr)$_z$.

In order to compare between the SPS and CVS techniques for the production of sintered SmCo$_5$ magnets, Fig. 3 shows the demagnetization curves at room temperature and 180 $^o$C for the SPS-ed sample produced at 900 $^o$C, as it showed to have the best coercivity among other SPS-ed samples, and a fresh SmCo$_5$ sample produced by CVS route. Table 2 shows a comparison between their density, oxygen content, $B_r$, $BH_{(max)}$, $_jH_c$ and external coercivity ($_bH_c$). At room temperature, the SPS-ed sample showed to have better $B_r$, coercivity and $BH_{(max)}$ in comparison to the CVS-ed samples, even though, the latter has better density and lower oxygen content than the former. This is likely to be due to the microstructure refinement after HD due to the large particle size distribution of the decrepitated powder [Eldosouky 2018]. Heating to 180 $^o$C has resulted in a more deterioration effect for the magnetic properties of the SPS-ed sample because of its lower density and the higher affinity for the interstitial oxidation. Nevertheless, even at 180 $^o$C, the latter is still showing good properties: $B_r$ of 0.436 T, $_jH_c$ of 1502 kA/m and $BH_{(max)}$ of 36.4 kJ/m$^3$, which confirms the applicability of the magnet for high temperature applications as in electric vehicles and space applications.

**FIG. 3 HERE**
**Table 2 HERE**

Fig. 4 shows the back scattered electrons (BSE) images of the CVS-ed sample and SPS-ed samples sintered at 800 $^o$C, 900 $^o$C and 1000 $^o$C, Fig. 5 shows their XRD patterns. The detected phases in the samples by the EDX were as follows: the matrix SmCo$_5$, randomly distributed light gray Sm$_2$Co$_7$ phase with random size, and white small spots which were analyzed to be a mixture of Sm-oxides and Sm-rich phases. From the XRD patterns, both the SPS-ed samples at 800 $^o$C and 900 $^o$C showed the presence of small fractions of the Sm$_2$Co$_{17}$ phase, with higher concentration in the 800 $^o$C samples. The appearance of Sm$_2$Co$_{17}$ is due to the increase of the oxygen concentration and the consumption of Sm in these samples and results in the decrease in the $_iH_c$ values for the samples sintered at temperature lower than 900 $^o$C, as mentioned above. The Sm$_2$Co$_{17}$ phase has disappeared by increasing the sintering temperature with increasing the percentage of the Sm$_2$Co$_7$ phase, where the SPS-ed samples at 1000 $^o$C showed to have Sm$_2$Co$_7$ with higher percentage than other samples, without the appearance of any traces of the Sm$_2$Co$_{17}$ phase.

**FIG. 4 HERE**
**FIG. 5 HERE**

## IV. CONCLUSION

SPS technique was successfully used for the densification of SmCo$_5$ powder produced by the HD of SmCo$_5$ sintered magnets at 4 bar. The properties of the SPS-ed samples showed to drastically depend on the sintering temperature. The SPS-ed sample produced by sintering at 900 $^o$C showed the most promising results among other SPS-ed samples, with $_jH_c$ higher than 1500 kA/m. The samples prepared by SPS route showed to have comparable results in comparison to fresh SmCo$_5$ samples made by CVS route. The SPS-ed samples also showed relatively good magnetic properties at 180 $^o$C. This indicates the possibility to use SPS technique for the production of SmCo$_5$ magnets of high coercivity which enable them to be used in various applications.

## ACKNOWLEDGMENT

The research leading to these results has received funding from the European Community's Horizon 2020 Programme ([H2020/2014-2019]) under Grant Agreement no. 674973 (MSCA-ETN DEMETER). This publication reflects only the authors' view, exempting the Community from any liability. Project website: http://etn-demeter.eu/. The authors thank Dr. Matejka Podlogar for the help with the microstructure analysis.

## REFERENCES

Diouf S, Molinari A (2012), "Densification mechanisms in spark plasma sintering: Effect of particle size and pressure," *Powder Technol.*, vol. 221, pp. 220–227, doi: 10.1016/j.powtec.2012.01.005.

Eldosouky A, Škulj I (2018), "Recycling of SmCo$_5$ magnets by HD process," *J. Magn. Magn. Mater.*, vol. 454, pp. 249-253, doi: 10.1016/j.jmmm.2018.01.064.

Fang L, Zhang T, Wang H, Jiang C, Liu J (2018), "Effect of ball milling process on coercivity of nanocrystalline SmCo$_5$ magnets" *J. Magn. Magn. Mater.*, vol. 446, pp. 200-205, doi: 10.1016/j.jmmm.2017.09.012.

Gutfleisch O, Willard M A, Brück E, Chen C H, Sankar S G, Liu J P (2011), "Magnetic materials and devices for the 21st century: stronger, lighter, and more energy efficient," *Adv. Mater.*, vol. 23, pp. 821–842, doi: 10.1002/adma.201002180.

European Commission (2017), "Study on the review of the list of critical raw materials," *EU Law and Publications*, pp. 1-93, doi: 10.2873/876644.

Harris I R (1987), "The potential of hydrogen in permanent magnet production," *J. Less-common Met.*, vol. 131, pp. 245–262, doi: 10.1016/0022-5088(87)90524-8.

Liu W Q, Cui Z Z, Yi X F, Yue M, Jiang Y B, Zhang D T, Zhang J X, Liu X B (2010), "Structure and magnetic properties of magnetically isotropic and anisotropic Nd-Fe-B permanent magnets prepared by spark plasma sintering technology," *J. Appl. Phys.*, vol. 107, pp. 09A719 (1-3), doi: 10.1063/1.3339067.

Lu Q M, Zhou C, Yue M (2015), "Enhanced magnetic properties and thermal stability of Nd$_2$Fe$_{14}$B/SmCo$_5$ composite permanent magnets prepared by spark plasma

sintering," *IEEE Trans. Magn.,* vol. 51(11), pp. 1-4, doi**:** 10.1109/TMAG.2015.2440477.

Mackie A J, Hatton G D, Hamilton H G C, Dean J S, Goodall R (2016), "Carbon uptake and distribution in spark plasma sintering (SPS) processed Sm(Co,Fe,Cu,Zr)$_z$," *Mater. Lett.,* vol. 171, pp. 14-17, doi: 10.1016/j.matlet.2016.02.049.

Menushenkov V (2006), "Phase transformation-induced coercivity mechanism in rare earth sintered magnets," *J. Appl. Phys.* vol. 99, 08B523 (1-3), doi: 10.1063/1.2176183.

Mo w, Zhang L, Shan A, Cao L, Wu J, Komuro M (2007), "Microstructure and magnetic properties of NdFeB magnet prepared by spark plasma sintering," *Intermetallics.,* vol. 15, pp.1483-1488, doi: 10.1016/j.intermet.2007.05.011.

Pan S (2014), *Rare earth permanent-magnet alloys high temperature phase transformation: in situ and dynamic observation and its application in material design.* Springer, pp. 1-7, doi: 10.1007/978-3-642-36388-7.

Saito T, Nishio-Hamane D (2018), "High-coercivity SmCo$_5$/α-Fe nanocomposite magnets," *J Alloys Compd.,* vol. 735, pp. 218-223, doi: 10.1016/j.jallcom.2017.11.060.

Saravanan P, Gopalan R, Sivaprahasam D, Chandrasekaran V (2013), "Intermetallics effect of sintering temperature on the structure and magnetic properties of SmCo$_5$/Fe nanocomposite magnets prepared by spark plasma sintering," *Intermetallics.,* vol. 42, pp. 199-204, doi: 10.1016/j.intermet.2013.05.011.

Walton A, Yi H, Rowson N A, Speight J D, Mann V S J, Sheriden R S, Bradshaw A, Harris I R, Williams A J (2015), "The use of hydrogen to separate and recycle neodymium-iron-boron-type magnets from electronic waste," *J. Clean. Prod.,* vol. 104, pp. 236-241, doi: 10.1016/j.jclepro.2015.05.033.

Yue M, Zuo J H, Liu W Q, Lv W C, Zhang D T, Zhang J X, Guo Z H, Li W (2011), "Magnetic anisotropy in bulk nanocrystalline SmCo$_5$ permanent magnet prepared by hot deformation," *J. Appl. Phys.,* vol. 109, pp. 07A711 (1-3), doi: 10.1063/1.3553933.

Zhang D T, Lv W C, Yue M, Yang J J, Liu W Q, Zhang J X, Qiang Y (2010), "Nanocrystalline SmCo$_5$ magnet synthesized by spark plasma sintering*,*" *J. Appl. Phys.,* vol. 107, pp. 09A701(1-3), doi: 10.1063/1.3334458.

Zhaohui Z, Fuchi W, Lin W, Shukui L, Osamu S (2008), "Sintering mechanism of large-scale ultra fine-grained copper prepared by SPS method," *Mater. Lett.,* vol. 62, pp. 3987–3990, doi: 10.1016/j.matlet.2008.05.036.

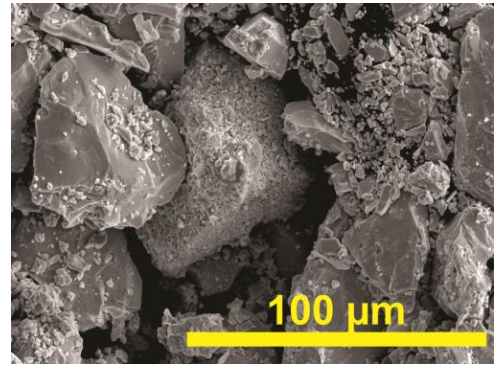

Fig. 1. SE image of the powder produced by the decrepitation of SmCo$_5$ magnets.

Table 1. Density, oxygen content and magnetic properties of the SPS-ed samples sintered at 800-1000 $^o$C.

| Temperature | $\rho$ [g/cm$^3$] | Oxygen cont. [wt%] | (BH)$_{max}$ [kJ/m$^3$] | $_jH_c$ [kA/m] | $B_r$ [T] |
|---|---|---|---|---|---|
| 800 $^o$C | 6.86 | 1.20 | 31.9 | 978.6 | 0.41 |
| 850 $^o$C | 7.14 | 0.76 | 30.5 | 960.0 | 0.40 |
| 900 $^o$C | 8.11 | 0.66 | 43.4 | >1500 | 0.47 |
| 950 $^o$C | 8.38 | 0.49 | 49.8 | 1243 | 0.51 |
| 1000 $^o$C | 8.40 | 0.47 | 50.1 | 1164 | 0.51 |

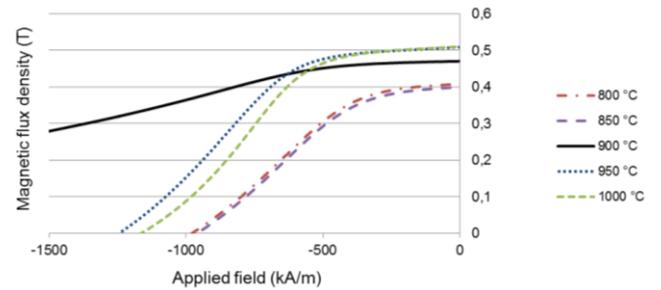

Fig. 2. The demagnetization curves for the SPS-ed samples sintered at 800 – 1000 $^o$C. The sample sintered at 900 $^o$C showed the best coercivity.

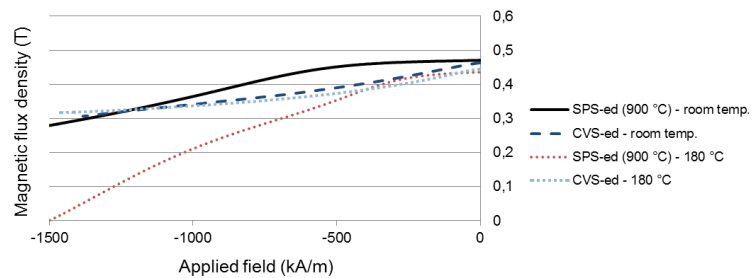

Fig. 3. Comparison between the demagnetization curves of the SPS-ed samples sintered at 900 $^o$C and the CVS-ed sample; at room temperature and 180 $^o$C.

Table 2. Density, oxygen content and magnetic properties of the SPS-ed samples sintered at 900 $^o$C and CVS-ed sample; at room temperature and 180 $^o$C.

| | (BH)$_{max}$ [kJ/m$^3$] | $_jH_c$ [kA/m] | $_bH_c$ [kA/m] | $B_r$ [T] | $\rho$ [g/cm$^3$] | Oxygen cont. [wt%] |
|---|---|---|---|---|---|---|
| SPS-ed (900 $^o$C) - room temp. | 43.4 | >1500 | 366.8 | 0.47 | 8.11 | 0.66 |
| CVS-ed - room temp. | 38.0 | >1500 | 328.4 | 0.46 | 8.25 | 0.4 |
| SPS-ed (900 $^o$C) -180 $^o$C | 36.4 | 1502 | 322.4 | 0.44 | - | - |
| CVS-ed - 180 $^o$C | 34.8 | >1500 | 314.2 | 0.44 | - | - |

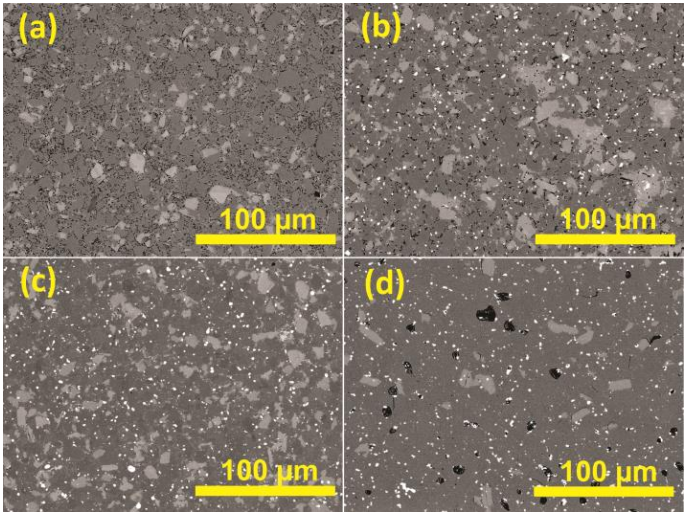

Fig. 4. The BSE images of the -ed samples sintered at 800 °C (a), 900 °C (b) and 1000 °C (c), and the CVS-ed sample (d).

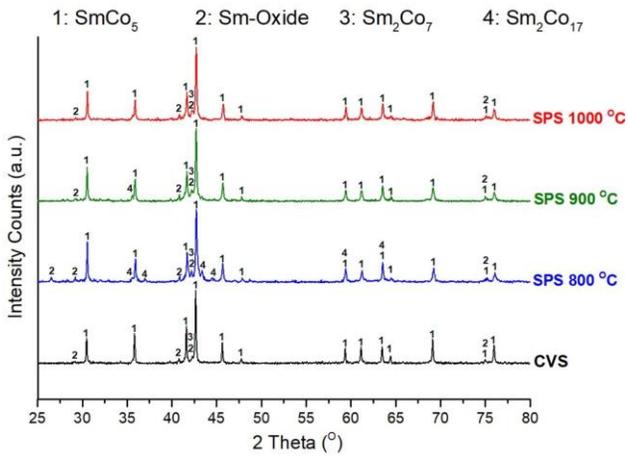

Fig. 5. The XRD patterns of the CVS-ed sample and SPS-ed samples sintered at 800 °C, 900 °C, and 1000 °C.